# Martingales, the Efficient Market Hypothesis, and Spurious Stylized Facts

Joseph L. McCauley, Kevin E. Bassler[+], and Gemunu H. Gunaratne[++]
Physics Department
University of Houston
Houston, Tx. 77204
jmccauley@uh.edu

[+]Texas Center for Superconductivity
University of Houston
Houston, Texas

[++]Institute of Fundamental Studies
Kandy, Sri Lanka

## Abstract

The condition for stationary increments, not scaling, detemines long time pair autocorrelations. An incorrect assumption of stationary increments generates spurious stylized facts, fat tails and a Hurst exponent $H_s=1/2$, when the increments are nonstationary, as they are in FX markets. The nonstationarity arises from systematic uneveness in noise traders' behavior. Spurious results arise mathematically from using a log *increment* with a 'sliding window'. We explain why a hard to beat market demands martingale dynamics , and martingales with nonlinear variance generate *nonstationary* increments. The nonstationarity is exhibited directly for Euro/Dollar FX data. We observe that the Hurst exponent $H_s$ generated by the using the sliding window technique on a time series plays the same role as does Mandelbrot's Joseph exponent. Finally, Mandelbrot originally assumed that the 'badly behaved second moment of cotton returns is due to fat tails, but that nonconvergent behavior is instead direct evidence for nonstationary increments. Summarizing, the evidence for scaling and fat tails as the basis for econophysics and financial economics is provided neither by FX markets nor by cotton price data.

## 1. Introduction

The finance and physics literature contains many papers claiming scaling via Hurst exponents on the one hand, and fat tailed distributions on the other. We can identify the 'central dogma of econophysics' as the expectations of Hurst exponent scaling, fat tailed distributions, and exponent universality. The question what are the underlying market dynamics has remained controversial. Some data analyses have indicated Levy models, others have suggested diffusive behavior. A martingale is a less restrictive class of diffusive model than a Markov process. The latter admits no memory, the former does.

The inference of Markov dynamics from empirical data would seem to be a reasonable first approximation because normal financial markets have finite variance and are very hard to beat. A Markovian market would be imposssible to beat. Finance markets appear to be Markovian to lowest order, but may contain exploitable memory at higher order. We define below precisely what we mean by 'lowest order' and why this suggests a martingale in log returns.

This paper is based on our new theoretical analysis [1] which was originally motivated by our recent foreign exchange (FX) data analysis [2]. The 'central dogma of econophysics', Hurst exponent scaling, universality, and fat tails is not exhibited by FX markets when the nonstaionarity of the increments is accounted for. Correspondingly, we illustrate explicitly why most existing data analysis claiming fat tails and scaling are wrong.

The analysis of this paper can be understood as a tale told by two different variables: First, there is what *we* define [1,2] as the **log return**

$$x(t) = \ln p(t)/p_c(t) \quad (1)$$

where p(t) is the price of a stock, bond, or foreign exchange at time t, and $p_c(t)$ can be understood as 'value' [3], the most probable price, the price that locates the peak of the 1-point returns density $f_1(x,t)$ at time t. Then, there is what most other theorists (beginning with Osborne) mean by log return,

$$x(t,T) = x(t+T) - x(t) = \ln p(t+T)/p(t), \quad (2)$$

but which is clearly an *increment* of the log return. We will explain that the log return x(t) is always a 'good' variable, both in theory and data analysis, and then we will explain why the use of the log increment x(t,T) in data analysis leads to spurious stylized facts, to spurious scaling with exponent $H_s=1/2$ and spurious fat tails in a wrongly extracted 1-point returns density $f_s$, where the subscript "s" denotes 'sliding window'. The two variables yield identical results iff. a data set or model generates stationary increments x(t,T)=x(T). We will show that the 1-point returns density $f_1(x,t)$ correctly extracted from FX market time series gives evidence neither for scaling with H over a time scale of a day, nor for fat tails. We speculate that stock prices, in contrast, may perhaps exhibit fat tails (but not Hurst exponent scaling) over the same time scale (there is no evidence for universality, and in far from equilibrium dynamics there is no reason to expect universality either).

Drift-free Markov, and more generally martingale processes generate uncorrelated increments that are generally nonstationary: $<x(t,T)x(t,-T)>=0$ where $x(t,-T)=x(t)-x(t-T)$. If the mean square fluctuation $<x^2(t,T)>$ depends on the starting time t then the increments are nonstationary. In FX (and in most other) data analyses stationary increments have been *implicitly* assumed by the use of a technique called a "sliding window". A 'sliding window' is used to build histograms by reading a time series while varying t in $x(t,T)$ with T fixed, and in the presence of nonstationary increments this method cannot generate $f_1(x,t)$. Instead, the method at best generates a spurious density $f_s(z,t,T)$ that we will define precisely below. The sliding window technique would be legitimate, would yield $f_1(z,T)$ independent of starting time t iff. the increments were stationary, iff. $z=x(t,T)=x(T)$ independent of the starting time t. But the increments in finance data are *not* stationary [2], and there is no ergodicity in a nonstationary (i.e., far from statistical equilibrium) time series, so that the sliding window method produces 'significant artifacts', spurious stylized facts. We emphaize that FX markets are a nonstationary stochastic process with nonstationary increments. All assumptions of stationarity fail miserably whe it comes to market data.

Another conclusion is that scaling doesn't matter anyway, *scaling gives us no information whatsoever about either the underlying market dynamics or memory*. The purpose of this paper is to explain all of these assertions, and to indicate how correctly to analyze random time series without generating spurious stylized facts. Our method and conclusions are not restricted to finance data but have application to the analysis of stocastically generated time series, whether in physics, economics, biology, or elsewhere. We offer a completely new viewpoint in the theory of stochastic processes and in data analysis.

## 2. Hurst exponent scaling

We define selfsimilarity of a stochastic process starting with the mathematicians' standpoint [4,5] and then show that their definition is equivalent to our definition [6,7] in terms of densities so long as the moments of the 1-point density are finite.

A stochastic process x(t) is said be selfsimilar with scaling exponent H, 0<H<1, if [4,5]

$$x(t) = t^H x(1), \quad (3)$$

where by equality we mean equality 'in distribution'. Note first that scaling trajectories necessarily pass through the 'filter' x(0)=0, trajectories with x(0)≠0 cannot possibly scale. Second, a method designed by Hurst to detect trends was originally used to define a different scaling exponent that Mandelbrot and Taqqu labeled the Joseph exponent J [8]. However, the notation "H" was used by Mandelbrot and van Ness [5] to describe fractional Brownian motion (fBm), a selfsimilar process that *does* produce the trends of the Hurst-Mandelbrot 'Joseph Effect' via long time pair correlations arising from stationarity of the increments. Mandelbrot and Taqqu distinguished H from J on the basis of Hurst's (highly nontransparent) R/S analysis, and noted that while H=J for fBm, for processes without long time increment autocorrelations, like Levy processes and drift free Markov processes, H≠J=1/2. Embrechts [4] denotes the selfsimilarity exponent by H, but stops short of writing H(urst). Because of the vast confusion in the scientific literature, wherein H≠1/2 is too often but wrongly thought to imply long time pair correlations, we will call the selfsimilarity exponent H "the Hurst exponent" and explain that selfsimilarity, taken alone, does not and cannot generate long time pair autocorrelations like those of fBm. We will introduce a second scaling exponent, the 'sliding window Hurst exponent' $H_s$, and will see that $H_s$ plays essentially the same role as does the Joseph exponent: $H \neq H_s = 1/2$ whenever there is selfsimilarity without long time pair correlations, but $H = H_s \neq 1/2$ in the presence of long time pair correlations combined with selfsimilarity. *The essential requirement for long time pair correlations will be seen to be stationarity of the increments with variance nonlinear in t, not selfsimilarity.* Selfsimilarity and stationarity of the increments are confused together into an unhealthy and misleading soup too often in the literature (see. e.g., the definition of the Hurst exponent in Wikipedia, http://en.wikipedia.org/wiki/Hurst_exponent, where the note added Oct., 2007, is ours). Next, we define selfsimilarity in terms of probability densities, which explains what is meant by asserting that $x(t) = t^H x(1)$ 'in distribution'.

The 1-point density $f_1(x,t)$ reflects the statistics collected from many different runs of the time evolution of $x(t)$ from a specified initial condition $x(t_o)$, where $x(0) = 0$ is required for scaling, but cannot describe correlations or the lack of same. Given a dynamical variable $A(x,t)$, the absolute (as opposed to conditional) average of A is

$$\langle A(t) \rangle = \int_{-\infty}^{\infty} A(x,t) f_1(x,t) dx \quad . \quad (4)$$

From (1), the moments of x must obey

$$\langle x^n(t) \rangle = t^{nH} \langle x^n(1) \rangle = c_n t^{nH} \quad (5)$$

Combining this with

$$\langle x^n(t) \rangle = \int x^n f_1(x,t) dx \quad (6)$$

we obtain [2]

$$f_1(x,t) = t^{-H}F(u), \quad (7)$$

where the scaling variable is $u=x/t^H$ [6].

In contrast, conditional averages $<A(x,t)>_{cond}$ require the 2-point density

$$f_2(x,t+T;y,t) = p_2(x,t+T|y,t)f_1(y,t), \quad (8)$$

or, more to the point, the 2-point transition density (conditional probability density) $p_2(x,t+T:y,t)$ [1].

If the absolute average of x(t) vanishes, then the variance is simply

$$\sigma^2 = \langle x^2(t) \rangle = \langle x^2(1) \rangle t^{2H}. \quad (9)$$

This explains what is meant by Hurst exponent scaling, and also specifies what's meant by asserting that eqn. (1) holds 'in distribution'. The vanishing of the absolute average of x does not mean that there's no conditional trend: in fractional Brownian motion (fBm), e.g., where $<x(t)>=0$ by construction, the conditional average of x does not vanish and depends on t [7], reflecting either a trend or anti-trend. In a Markov process, scaling requires that the drift rate depends at worst on t (is independent of x) and has been subtracted, that by "x" we really mean the detrended variable $x(t)-\int R(s)ds$. Markov processes with x-independent drift can be detrended over a definite time scale, but any attempt to detrend fBm is an illusion because the 'trend' is due to long time autocorrelations, not to an additive drift term [1] that can be removed. The attempt to detrend a time series implicitly assumes an underlying martingale plus drift, and fBm is by construction not a martingale plus drift.

Hurst exponent scaling is confined to 1-point densities and simple averages, and *1-point densities cannot be used to identify the underlying stochastic dynamics* [1,9,10]. Even if scaling holds at the 1-point level as in fBm, the 2-point density (the transition density $p_2$) and the pair correlations $<x(t+T)x(t)>$ do *not* scale, and it's the transition density $p_2$ that's required to give a minimal description of the underlying dynamics[1]. *In particular, scaling, taken alone, implies neither the presence nor absence of autocorrelations in increments/displacements taken over nonoverlapping time intervals.* That is, scaling has nothing whatsoever to say about whether a market is effectively efficient (hard to beat), or is easily beatable, in contrast to what one of us incorrectly assumed earlier [3,11].

The financial economics literature is filled with wrong claims and wrong assumptions about financial time series. These misconceptions are systematically and hopelessly propagated whenever a researcher uses the standard statistical methods of econometrics. In Fama [12], e.g., the claim is made that returns are uncorrelated, $<x(t+T)x(t)>=0$. More recently, in Lux and Heitger [13] the statement is made that prices are random (by 'random' we understand 'uncorrelated') but returns are not random. The correct statements, explained below, are that both prices and returns are always correlated, $<p(t+T)p(t)>\neq 0$, $<x(t+T)x(t)>\neq 0$, but *increments* in returns approximately vanish after a trading time of 10 minutes: $<x(t,T)x(t,-T)>\approx 0$ for T≥10 min. of trading [2]. The latter is basically the effectively efficient market hypothesis: one cannot make money systematically by trading on either simple averages or pair correlations [1]. Note that an assumption of stationary increments, the confusion of x(t,T) with x(T), would lead one wrongly to assert that returns are uncorrelated.

---
[1] For a Gaussian process, pair correlations and $p_2$ provide a complete description. But for nonGaussian processes like FX markets all of the transition densities $p_n$, n=2,3,… are required to pin down the dynamics. In practice, we usually do not know any more about the dynamics than can be extracted from pair correlations.

## 4. Stationary vs. nonstationary increments

*Stationary processes* are often confused with *stationary increments* in the literature (see [6] for a discussion). Stationary increments are implicitly assumed in data analyses and simulations whenever a sliding window method is used to extract histograms, and the sliding window method is most commonly found in the literature [15,16,17,18]. We define stationary and nonstationary increments and exhibit their implications for the question of long time autocorrelations, or complete lack of autocorrelations. We show that the question of stationary increments, not scaling, is central for the existence of long time correlations.

By increments, we mean displacements x(t,T)=x(t+T)-x(t). Stationary increments of a nonstationary process x(t) are defined by [4]

$$x(t + T) - x(t) = x(T), \qquad (10)$$

and by nonstationary increments [1,7,14] we mean that the difference

$$x(t + T) - x(t) \neq x(T) \qquad (11)$$

depends on both (t,T), not on T alone. The implications of this distinction for data analysis, and for understanding Hurst exponents, are central. In the nonstationary case the density of increments z=x(t,T) must be obtained from the two-point density $f_2(x(t),x(t+T);t,t+T)$ and depends on starting time t,

$$f_s(z,t,T) = \int dxdy f_2(y,t+T;x,t)\delta(z-y+x), \quad (12)$$

where the subscript s here denotes 'sliding window', whereas for stationary increments this 1-point density is independent of t, $f_s(x,t,T)=f_s(z,T)$. In the latter case, the density $f_s$ is not spurious and sliding windows can be used to extract correct histograms reflecting $f_1(z,T)$ from a single long time series.

The efficient market hypothesis (EMH) [15] is sometimes interpreted to mean that the market is impossible to beat, that there are no correlations at all (no systematically repeated price/returns patterns) that can be exploited for profit. A Markov market satisfies the condition of an impossible to beat market. Because real markets are very hard (if not necessarily impossible) to beat, models that generate no autocorrelations in increments are a good zeroth order approximation to real markets [1]. In such models, the autocorrelations in increments x(t,T) and x(t,-T) vanish

$$\langle (x(t_1) - x(t_1 - T_1))(x(t_2 + T_2) - x(t_2)) \rangle = 0, \quad (13)$$

if there is no time interval overlap,

$$[t_1 - T_1, t_1] \cap [t_2, t_2 + T_2] = \Phi, \quad (14)$$

where $\Phi$ denotes the empty set on the line. This is a much weaker and more pregnant condition than would be asserting that the increments are statistically independent. Condition (14) is in fact a martingale condition in weak disguise. Eqn. (14) means that nothing that happened in an earlier time interval can be used to predict systematically the returns in a later time interval *at the level of (simple averages*

*and) pair correlations*. That is, the market is 'effectively efficient' in the sense that simple averages and pair correlations look Markovian, unlike fBm there is no memory in pair correlations to be exploited. This may not rule out higher order correlations that might be used for technical trading. I.e., a Markovian market is 'efficient' in the strictest sense, is impossible to beat, whereas a martingale market looks Markovian to lowest order (at the level of simple averages and pair correlations), but might be systematically beatable at some higher level of insight. This defines precisely what we mean by "lowest order".

Consider a stochastic process x(t) where the increments are uncorrelated. From this condition we easily obtain the autocorrelation function for returns x(t)

$$\langle x(t)x(s) \rangle = \langle (x(t) - x(s))x(s) \rangle + \langle x^2(s) \rangle = \langle x^2(s) \rangle > 0, \quad (15)$$

since $x(s)-x(t_o)=x(s)$, so that $<x(s)x(t)>=<x^2(s)>=\sigma^2$ is simply the variance in x. *This is a martingale condition*,

$$\langle x(t+T) \rangle_{cond} = x(t), \quad (16)$$

or

$$\int dy\, y\, p_2(y, t+T | x, t) = x. \quad (17)$$

The result has a nice interpretation: since $<x(t,T)x(s)>=0$ for $s \leq t < t+T$, future 'gains' $x(t,T)$ are uncorrelated with all past returns.

We next obtain another central result. Combining

$$\langle(x(t+T)-x(t))^2\rangle = +\langle x^2(t+T)\rangle + \langle x^2(t)\rangle - 2\langle x(t+T)x(t)\rangle$$
(18)

with (14), we get

$$\langle(x(t+T)-x(t))^2\rangle = \langle x^2(t+T)\rangle - \langle x^2(t)\rangle \quad (19)$$

which depends on both t and T, excepting the rare case where the variance $\langle x^2(t)\rangle$ is linear in t. *Martingale increments are uncorrelated and are generally nonstationary.* I.e., we must expect nonstationary increments in effectively efficient markets. The variance $\langle x^2(t)\rangle$ of a real FX market is not linear in t, it has instead very complicated variation with time.

Consider next the class of all stochastic processes with stationary increments, x(t,T)=x(T) 'in distribution'. Here, we begin with

$$-2\langle x(t+T)x(t)\rangle = \langle(x(t+T)-x(t))^2\rangle - \langle(x^2(t+T)\rangle - \langle x^2(t)\rangle,$$
(20)

and then using (8) on the right hand side of (18) we obtain

$$-2\langle x(t+T)x(t)\rangle = \langle(x^2(T)\rangle - \langle(x^2(t+T)\rangle - \langle x^2(t)\rangle \quad (21)$$

which differs from (13). The increment autocorrelation function is

$$2\langle(x(t)-x(t-T))(x(t+T)-x(t))\rangle = \langle x^2(2T)\rangle - 2\langle x^2(T)\rangle \quad (22)$$

which vanishes iff. the variance $\langle(x^2(t)\rangle$ is linear in t. *Stationary increments are typically strongly correlated.*

E.g., if scaling (1) holds then we obtain the prediction of infinitely long time autocorrelations

$$\langle (x(t) - x(t-T))(x(t+T) - x(t)) \rangle = \langle x^2(T) \rangle (2^{2H-1} - 1).$$
(23)

characteristic of fBm [5,7]. This autocorrelation vanishes iff. H=1/2, otherwise the autocorrelations are strong for all time scales T. Such fluctuations violate the EMH, especially if H cannot be approximated as H≈1/2. Note that scaling is not the essential point, is in fact irrelevant: *stationarity of the increments, reflected in the t-independent pair correlations (21), is the central requirement for long time increment autocorrelations.*

Summarizing, the Hurst exponent H tells us nothing whatsoever about autocorrelations in increments, tells us nothing whatsover about the underlying dynamics apart from scaling itself, and tells us nothing whatsoever about the effficiency or lack of same of a market. In the next two sections we will sharpen the distinction by exhibiting both scaling Markov processes and fBm where H≠1/2.

**Scaling Ito Processes**

An Ito process is generated locally by the stochastic diffferential equation (sde)

$$dx = R(x,t) + \sqrt{D(x,t)}dB(t) \quad (24)$$

where B(t) is the Wiener process. A Wiener process is an uncorrelated Gaussian process scaling with H=1/2, so that the increments are stationary and (from Ito's theorem) $dB^2 = dt = \langle dB^2 \rangle$. Iff. R(x,t)=R(t) is independent of x can we detrend all trajectories once and for all by replacing x(t) by x(t)-∫R(s)ds. With this substitution, the Ito process is a martingale. The absolute average gives <x(t)>=0 and there is no trend. Finite memory may be present but we will not write the possible memory explicitly. Instead,

The variance can be calculated from the stochastic integral of (24) as

$$\sigma^2 = \int_0^t ds \int_{-\infty}^{\infty} dx\, f_1(x,s) D(x,s), \qquad (25)$$

where x(0)=0, so that scaling of the density and the variance imply that the diffusion coefficient scales as well [6]:

$$D(x,t) = t^{2H-1} D(u), \quad u = x/t^H. \qquad (26)$$

Note that scaling of D does not imply scaling of the transition density $p_2(x,t+T;x_o,t)$.

We can also write the mean square fluctuation about an arbitrary point x(t) globally as

$$\langle (x(t+T)-x(t))^2 \rangle = \int_t^{t+T} ds \int_{-\infty}^{\infty} dx\, f_1(x,s) D(x,s) = \langle x^2(1) \rangle ((t+T)^{2H} - t^{2H})$$
(27)

and locally for t>>T as

$$\langle (x(t+T)-x(t))^2 \rangle \approx t^{2H-1} D(u) T. \qquad (28)$$

Both the global and local mean square fluctuations are useful in FX data analysis. In particular, in () the mean square fluctuation scales with T with $H_s=1/2$.

An Ito stochastic process may have finite memory. By 'finite memory' we mean a 'filtration' $(x_{n-1}, x_{n-2}, \ldots, x_1)$ that every trajectory must pass through. An example with n=2 is given in [16].

Ito processes are 1-1 with Fokker-Planck pdes [16] so we work with the drift free Fokker-Planck pde

$$\frac{\partial p_2}{\partial t} = \frac{1}{2}\frac{\partial^2}{\partial x^2}(Dp_2), \qquad (29)$$

where scaling may occur at best only for $f_1(x,t) = p_2(x,t:0,0,)$.

Model 1-point densities that scale with H are easily calculated [6,17,18]. With

$$f_1(x,t) = t^{-H}F(u); u = x/t^H \qquad (30)$$

and

$$D(x,t) = t^{2H-1}D(u), u = x/t^H \qquad (31)$$

the Fokker-Planck pde (32) yields

$$2H(uF(u))' + (D(u)F(u))'' = 0 \qquad (32)$$

which we integrate to obtain

$$F(u) = \frac{C}{D(u)}e^{-2H\int u du/D(u)} \qquad (33)$$

*For H≠1/2 all of these processes generate nonstationary increments.*

If

$$D(u) = (1+|u|)/2H \qquad (34)$$

Then we get the exponential density

$$F(u) = Ce^{-|u|}, \qquad (35)$$

where C is the normalization constant. For FX data a 2-sided exponential density is needed and is easily derived.

## 10. FX market facts vs. spurious stylized facts

We begin with 'the observed stylized facts' as understood by Holmes [29]: (i) asset prices are persistent and have, or are close to having, a unit root and are thus (close to) nonstationary; (ii) asset returns are fairly unpredictable, and typically have little or no autocorrelations; (iii) asset returns have fat tails and exhibit volatility clustering and long memory. Autocorrelations of squared returns and absolute returns are significantly positive, even at high-order lags, and decay slowly; (iv) Trading volume is persistent and there is positive cross-correlation between volatility and volume. These statements reflect a fairly standard set of expectations. Next, we contrast those expected stylized facts with the hard results of our recent FX data analysis [2]. Our analysis is based on 6 years of Euro/dollar exchange rates taken at 1 min. intervals.

The intended meaning of point (i) above is unclear because 'pesistence' is not defined, and a hard to beat market (an approximately efficient market) cannot exhibit persistence of the sort described by fBm. Furthermore, increment autocorrelations in FX market returns will vanish after about 10 min. of trading. Worse, a simple coordinate transformation $x(t)=\ln p(t)$ cannot erase persistence, whatever persistence might be. (ii) Both prices and returns have positive autocorrelation, $<x(t+T)x(t)>=<x^2(t)> > 0$, and autocorrelations in increments are approximately zero after 20 min. of trading, $<x(t,T)x(t,-T)>\approx 0$. It would appear that $x(t,T)$ has been confused with $x(T)$. (iii) We find no evidence for fat tails, and no evidence for Hurst exponent scaling on the time scale of a day. Because of nonstationarity of the increments, a 7 yr. FX time series is far too short (the histograms have too much scatter due to too few points) to indicate what may happen on larger time scales. Although we do not present the proof here, volatility clustering does not indicate 'long memory' but is explained as a purely Markovian phenomenon for variable diffusion processes, stochastic processes with diffusion coefficients $D(x,t)$ where the $(x,t)$ dependence is inherently nonseparable [20]. About point (iv) above, we offer no comment in this paper.

Our main point is: *the data analyses usedto arrive at the expected stylized facts have all used a technique called 'sliding windows'* [2]. The aim of this section is to explain that sliding windows produce spurious, results because FX data are nonstationary processes with nonstationary increments. Only one previous FX data analysis [21] that we are aware of showed that sliding windows lead to a spurious Hurst exponent $H_s=1/2$, and correctly identified the cause as nonstationarity of the increments. We explain that result theoretically below, and in addition have shown theoretically how sliding windows generate spurious fat tails [22] as well.

Here's what's meant by the sliding window method: one treats the increment $z=x(t,T)$ as if it would be indepedet of time of day t, and attempts to construct histograms $f_1(z,T)$ for increments at differentlag times T by reading a time series of returns x(t). There, one starts at initial time t and forms a window at time t+T. One assumes that the increment z=x(T,t)=x(t+T)-x(t) generates a 1-point density that is independent of t by sliding the window along the entire length of the time series, increasing t by one unit at a time while holding T fixed. For a long time series, one of at least $t_{max}$≈several years in length, this method is expected to produce good statistics because it picks up a lot of data points. **But the histograms generated from varying t in the increments x(t,T) yield $f_1(z,T)$ independently of t iff. the increments are stationary,** *otherwise the assumption is false.* And the assumption *is* false: first, we've shown [2] that the increments are uncorrelated after about 10 min. Second, fig. 1 shows that the mean square fluctuation $<x^2(t,T)>$ with T fixed at 10 min. depends very strongly on t throughtout the course of a trading day. This means simply that the traders' noisy behavior is not independent of time of day. *Our conclusion is that FX data, taken at 10 min. (or longer) intervals are described by a martingale with nonstationary increments in log returns.*

To illustrate how spurious stylized facts are generated by using a sliding window in data analysis, we apply that method to a time series with uncorrelated nonstationary increments, one with no fat tails and with a Hurst exponent H=.35, namely, a time series generated by the exponential density (16) with H=.35 (figure 2a) and linear diffusion (41). The process is Markovian. Fig. 1a was generated by taking 5,000,000 independent runs of the Ito process, each starting from x(0)=0 for T=10, 100, and 1000. The sliding window result is shown as figure 2b. In this case, the sliding windows appear to yield a scale free density $F_s(u_s)$, $u_s=x_s(T)/T^{Hs}$, from an empirical average over t that one cannot formulate theoretically, because for a nonstationary process there is no ergodic theorem. Not only are fat tails generated artificially, but we get a spurious Hurst exponent $H_S=1/2$ as well. *This is the method that has been used to generate stylized fact' in nearly all existing finance data analyses.*

Next, we describe our study of a six year time series of Euro-Dollar exchange rates from Olsen & Associates [2]. The increments $x(t,T)=x(t+T)-x(t)=\lm(p(t+T)/p(t))$ are nonstationary, *as is shown by the root mean square fluctuation in increments plotted against t in figure 1*, where T=10 min. to insure that there are no autocorrelations in increments. Second, note that the returns data do not scale with a Hurst exponent H or even with several different Hurst exponents over the course of a trading day (we define a trading day in a 24 hour market by resetting the clock at the same time each morning). We've shown [2] that the same stochastic process is repeated on different days of the week, so that we can assume a *single*, definite intraday stochastic process x(t) in intraday returns. In fig. 1 we see that scaling is observed at best within four disjoint time intervals during the day, and even then with four different Hurst exponents (H<1/2 in three of the intervals, H>1/2 in the other). *That is, the intraday stochastic process x(t) generally does not scale and will exhibit a complicated time dependence in the variance $<x^2(t)>$.*

Within the three windows where a data collapse $F(u)=t^H f(x,t)$ is weakly but inadequately indicated, we see that the scaling function F(u) has no fat tails, is instead approximately exponential (figure 3a). If we apply the method of sliding windows to the finance time series within the interval I shown in fig. 1, then we get figure 3b, which has artificially generated fat tails and also a spurious Hurst exponent $H_S=1/2$, just as with our numerical simulation using time series generated via the exponential density to generate a Markov time series (fig. 2a,b). *This shows how sliding windows can generate artificial fat tails and spurious Hurst*

*exponents of 1/2 in data analysis*. That is, the use of sliding windows generates 'spurious stylized facts' when the increments are nonstationary. This observation has far reaching consequences for the analysis of random time series, whether in physics, economics/finance, and biology.

Using the short time approximation T<<t, where t ranges from opening to closing time over a day, we obtain from (27) the mean square fluctuation

$$\langle x^2(t,T)\rangle \approx D(x,t)T = t^{2H-1}D(u)T \quad (36).$$

With uncorrelated nonstationary increments, in a scaling region we have more generally from (34) that

$$\langle x^2(t,T)\rangle = \langle (x(t+T)-x(t))^2\rangle = \langle x^2(1)\rangle[(t+T)^{2H}-t^{2H}}] \quad (37)$$

independent of the details of the diffusion coefficient D(x,t). In most existing data analyses we generally have T/t<<1 when sliding windows are applied to the increments x(T,t), yielding

$$\langle x^2(t,T)\rangle \approx \langle x^2(1)\rangle 2Ht^{2H-1}T. \quad (38)$$

Sliding windows then average empirically over t,

$$\langle x^2(t,T)\rangle_s \approx \langle x^2(1)\rangle 2H\langle t^{2H-1}\rangle_s T \quad (39)$$

yielding $<x^2(t,T)>_s \approx T^{2H_s}$ with $2H_s=1$. Sliding window Hurst exponents $H_s=1/2$ have been reported several times in the literature [34], but without any correct explanation how they arise from models where increments are uncorrelated with

H≠1/2. That $H_S$=1/2 is a consequence of using sliding windows was first reported by Galluccio et al [21] in 1997 in a paper that we did not appreciate at all until we rediscovered the implications of nonstationary increments for ourselves. In 1996 there was no theory available as guide.

Our exponent sliding window $H_s$ plays the same role for scaling martingales and fBm as does the Joseph exponent J: when there is scaling with H≠1/2 and with no increment autocorrelations then H≠$H_s$=1/2, whereas for stationary increments with nonlinear variance that scales with H then H=$H_s$. One need not use R/S analysis [8] to look for long time correlations, one need only check the mean square fluctuation <$x^2$(t,T)> for lack of t-dependence, for stationary increments.

Finally, consider figure 2 in Mandelbrot [23], where fat tails with infinite variance were deduced for cotton returns. He plots what he calls a 2$^{nd}$ moment, but which is analogous to the mean square fluctuation in our fig. 1 and is simply our eqn. (38) above. Mandelbrot observes that quantity is 'badly behaved', doesn't 'converge', and assumes without proof that the cause is Levy-like fat tails (in a Levy density the variance is strictly infinite). He then set the 2$^{nd}$ moment equal to infinity, *assuming that the time series is stationary* so that his sliding window time average (our eqn. (38)) should make sense. But markets are nonstationary, are very far from statistical equilibrium, and in that case an ergodicity assumption about the empirical time average in eqn. (38) fails, the mean square fluctuation in (38) will not 'converge' but will fluctuate eradically if the increments are nonstationary. ***The 'bad behavior' observed by Mandelbrot has nothing to do with fat tails and is instead direct evidence for nonstationarity of the increments.*** His figure 2 is nothing more or less than the uneveness exhibited by noise traders like our fig. 1. We have produced evidence that FX traders reproduce the same dynamics day after day, so the natural time scale for that analysis is one day. For cotton returns, the natural time scale for a correct data analysis is probably one year, with nonstationarity of increments reflecting unevenness of trading during the course of a year. Such seasonal variations cannot be smoothed out without masking the essence of the underlying market dynamics. It would be of interest to check cotton market returns for uncorrelated increments (to check for a martingale), where the diffusion coefficient (as explained above) would then describe the uneveness in the volatility of trading (the nonstationarity of the increments) over the time scale of a year. But a reliable cotton market analysis is made extremely difficult than FX because cotton price statistics are much more sparse, and will yield far more scatter in histograms than do FX market statistics where we cannot even get good enough daily returns histograms from 6 years of trading taken at 10 min. intervals. We would expect agricultural commodities in general to exhibit nonstationary increments.

## Acknowledgement

KEB is supported by the NSF through grant #DMR-0427938. GHG is supported by the NSF through grant #DMS-0607345 and by TLCC. JMC thanks Enrico Scalas and Harry Thomas for very helpful email discussions. This paper was presented by JLM as the evening lecture at the *Econophysics Colloquium and Beyond* in Ancona, Sept., 27-29, 2007. JLM is grateful to Mauro Gallegatti for the invitation to speak, and is also grateful to Edward Murphy for suggesting that we look at cotton.

**Figure Captions**

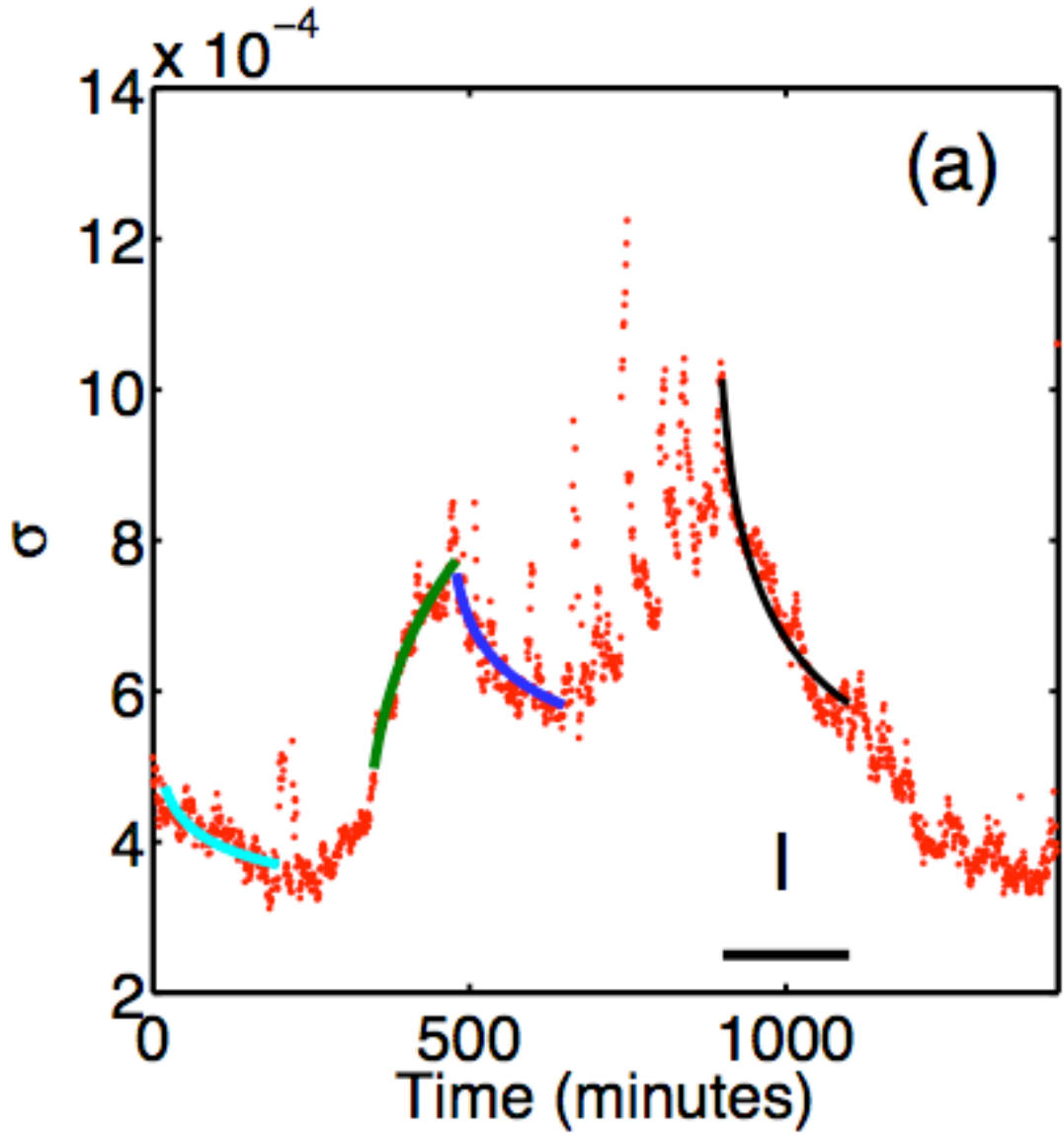

Fig. 1. The root mean square fluctuation $\langle x^2(t,T)\rangle^{1/2}$ of the daily Euro-Dollar exchange rate is plotted against time of day t, with T=10 min. to insure that autocorrelations in increments have died out (fig. 3).

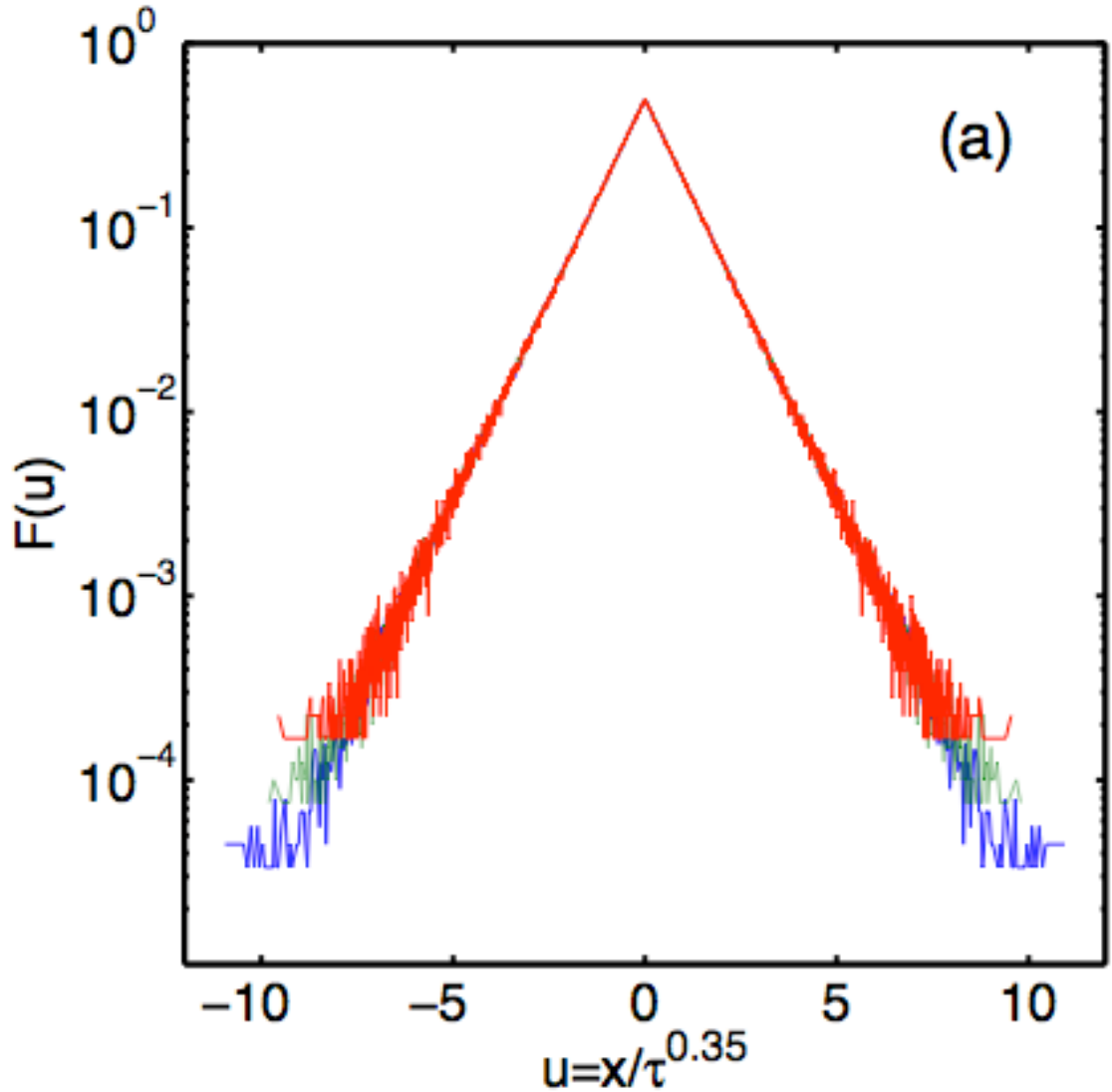

Fig. 2(a). The scaling function F(u) is calculated from a simulated time series generated via the exponential model, D(u)=1+abs(u) with H=.35. 5,000,000 independent runs of the exponential stochastic process were used.

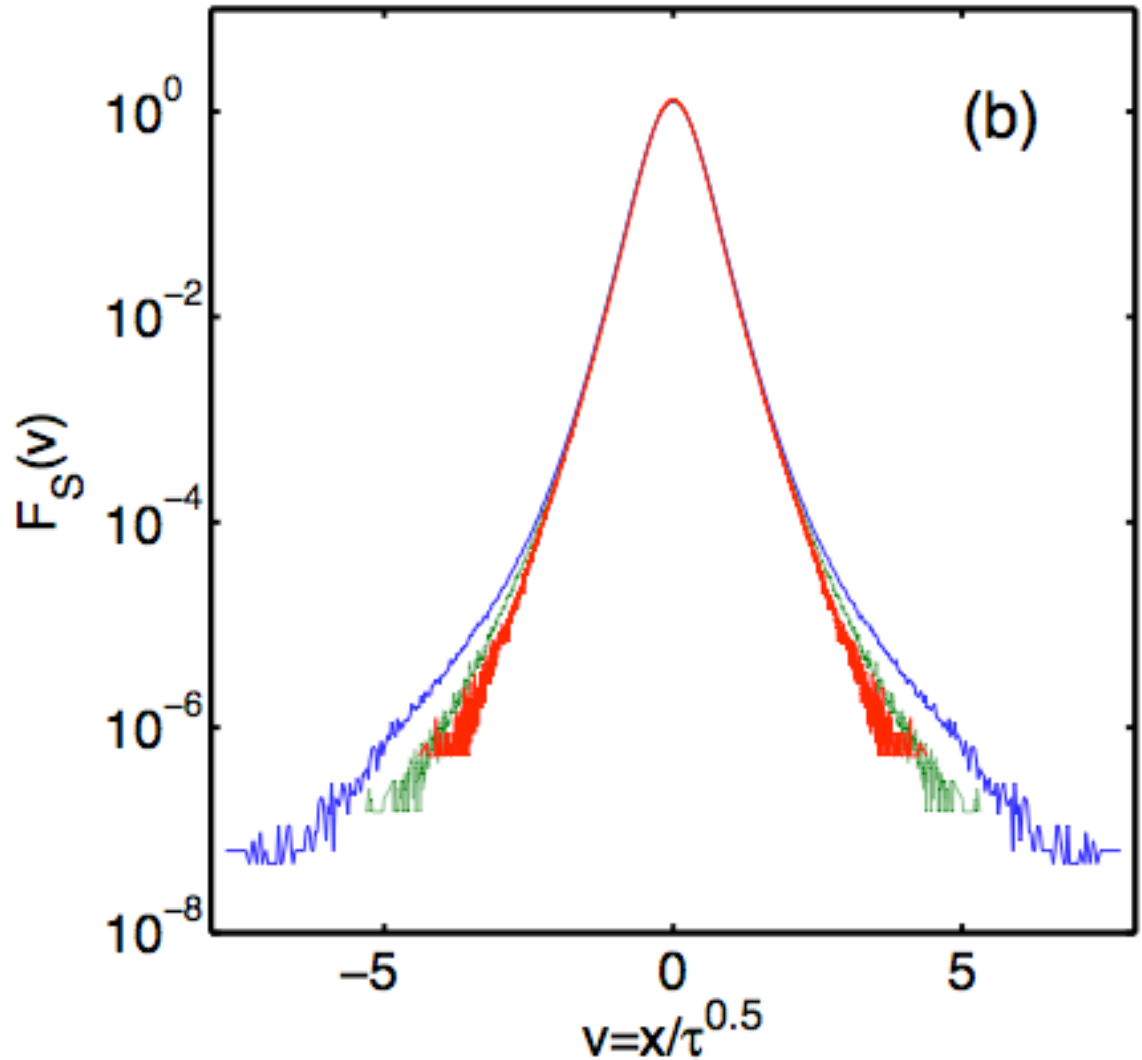

Fig. 2(b) The 'sliding window scaling function' $F_S(u_s)$, $u_s = x_s(T)/T^{H_s}$ was calculated for the same simulated data. Note that $F_S$ has fat tails whereas $F$ does not, and that $H_S=1/2$ aprears contradicting the fact that $H=.35$ was used to generate the time series. That is, sliding windows produce two significantly spurious results.

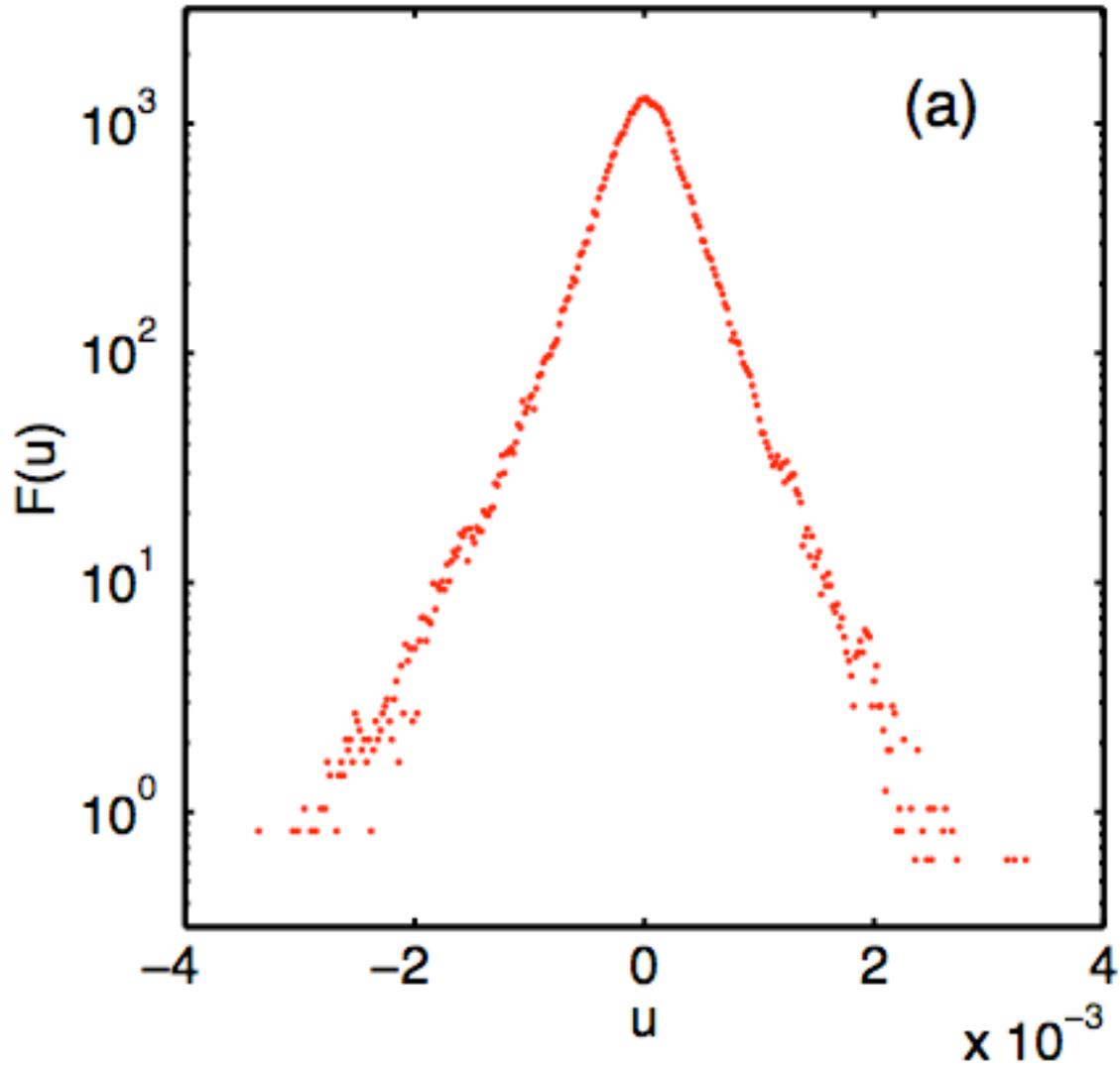

Fig. 3(a). Our scaling analysis uses the small window I shown in fig. 4a. We plot the scaling function F(u) for H=.35 with 10 min. ≤ T ≤ 160 min. Note that F(u) is slightly asymmetric and is approximately exponential, showing that the variance is finite.

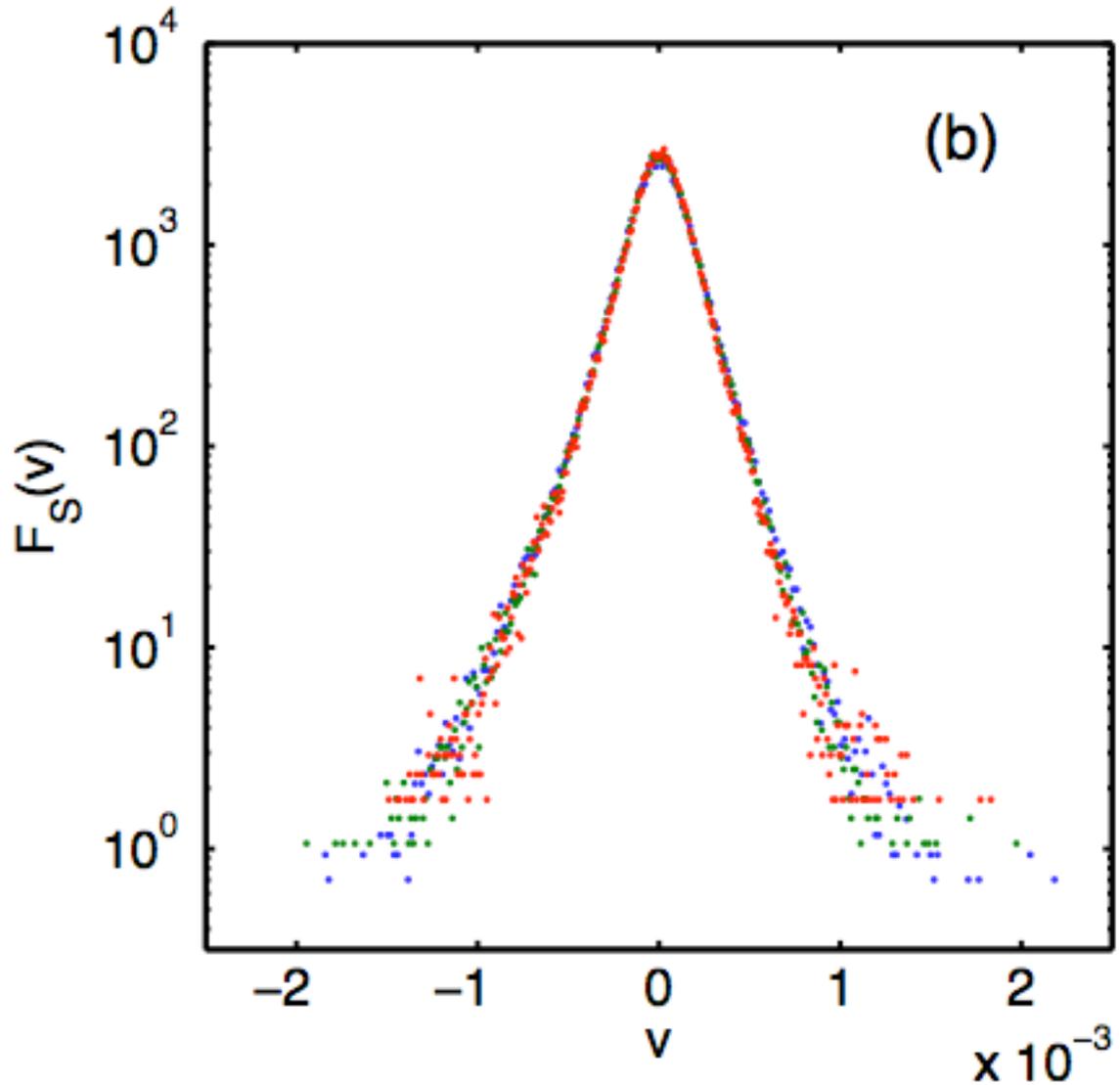

Fig. 3(b) The 'sliding interval scaling function' $F_s(u_s)$, $u_s = x_s(T)/T^{H_s}$, is constructed empirically from the same interval I for T=10, 20, and 40 min. Note that fat tails have been generated spuriously by the sliding windows, and that a spurious Hurst exponent $H_s=1/2$ has been generated as well, just as in the simulation data shown as fig. 3a,b.